\documentclass[aps,pra,twocolumn,showpacs]{revtex4-1}
\usepackage{amsmath,amsthm}  
\usepackage{epsf}
\usepackage{amssymb}  
\usepackage{graphicx}
\newcommand{\ket}[1]{| #1 \rangle}

\newcommand{\eq}[1]{Eq.~\eqref{#1}}

\begin{document}
\title{Thermal phase transitions for  Dicke-type models in the ultra-strong coupling limit}
\author{M.~Aparicio Alcalde $^{a}$, M.~Bucher$^{b}$,  C.~Emary $^{b}$ and T.~Brandes$^{b}$}
\affiliation{ 
$~^{a}$ 
Instituto de F\'{\i}sica Te\'orica, UNESP - S\~ao Paulo State University,
Caixa Postal 70532-2, 01156-970 S\~ao Paulo, SP, Brazil. \\
$^{b}$ Institut f\"ur Theoretische Physik,
  Hardenbergstr. 36,
  TU Berlin,
  D-10623 Berlin,
  Germany}
\date{\today{ }}
\begin{abstract}
We consider  the Dicke model in the ultra-strong coupling limit to investigate thermal phase transitions and their precursors at finite particle numbers $N$ for bosonic and fermionic systems. 
We derive partition functions with degeneracy factors that account for  the number of configurations and derive explicit expressions  
for the Landau free energy. This allows us to discuss the difference between  the original Dicke (fermionic) and the bosonic case. We find a crossover between these two cases that shows up, e.g., in the 
specific heat. 
\end{abstract}



\maketitle

The Dicke model, in its original weak-coupling and multi-mode form, \cite{Dic54},  has a long history as a paradigm for collective dissipation \cite{Bra05}, and, in its single mode form, as a test-bed for fundamental concepts such as the quantum-classical relation \cite{EB03two,AH12}, scaling \cite{VD06}, or entanglement \cite{LEB04} near quantum phase transitions. 
Interest in the Dicke superradiance model has been furthered by the recent discovery of the `Hepp-Lieb'-type  quantum phase transition \cite{HL73,PhysRevA.75.013804} with  Bose-Einstein condensates in an optical cavity \cite{BGBE10}. 
As a mean-field-type phase transition, the full phase diagram in the temperature-coupling constant plane was derived early 
\cite{WH73,HL73a,CGW73}
in the thermodynamic limit of $N\to \infty$ particles, cf. also \cite{ALS07}.
 In this Brief Report, we re-examine the {\em thermal} properties of (a somewhat generalized version of) this model with particular emphasis on the influence of the quantum statistics on the quantum phase transition.  We only consider
the ultra-strong coupling limit between the atoms and the light (corresponding to the superradiant phase at low temperatures), but we distinguish between various cases of $N$ bosons or $N$ fermions
distributed among $N_s$ two-level sites. In particular, we derive simple expressions for the thermodynamic partition sums that can be used to easily calculate thermodynamic quantities such as the specific heat at finite particle number $N$, and to follow an interesting crossover between the case of $N$ bosons on $N_s=1$ site and the original Dicke case of  $N$ fermions on $N_s=N$ sites .

As a starting point we use the Dicke Hamiltonian with the single bosonic mode $a, a^{\dagger}$ of frequency $\omega$. The angular momentum operators $J_{z}, J^{\pm}$ describe an ensemble of $N$ two-level atoms with a level splitting $\omega_{0}$. The single mode Dicke Hamiltonian is 
\begin{equation}\label{HDicke}
\mathcal{H}=\omega a^\dagger a + \omega_0 J_z + \frac{g}{\sqrt{N}}(a+a^\dagger)(J^++J^-).
\end{equation}
A unitary transformation with $U =e^{\sigma J_z} e^{i\frac{\pi}{2}J_y} $ and $\sigma\equiv \frac{2g}{\sqrt{N}\omega}(a^\dagger-a)$  rotates and polaron-transforms the Hamiltonian into $\mathcal{H}'\equiv U\mathcal{H}U^\dagger $ with
\begin{equation}
\mathcal{H}' = -\frac{\omega_0}{2}(J^+e^\sigma+J^-e^{-\sigma}) + \omega  a^\dagger a  -\frac{(2g)^2}{N\omega}J_z^2.
\end{equation}
The Hamiltonian $\mathcal{H}'$ can be used as a starting point for a perturbation theory in $\omega_0$, i.e. around the limit of very large coupling $g\to \infty$ between the angular momentum and the photon mode. In this limit, the physics is then determined by a (trivial) free photon   Hamiltonian $\omega  a^\dagger a$
and the angular momentum part $\propto J_z^2$. 
The analysis of the  properties of the thermodynamic partition sum 
\begin{equation}
Z_N\equiv {\rm Tr} e^{-\beta\mathcal{H}_N},\quad \mathcal{H}_N\equiv -\frac{(2g)^2}{N\omega}J_z^2
\label{z-general}
\end{equation} 
for different physical realizations of $J_z$ is the aim of this Brief Report.

The r\^{o}le of particle statistics in $Z_N$  can be qualitatively understood by considering the existence or otherwise of a thermal phase transition for $\mathcal{H}_N$ in the limit of $N\to \infty$. To this end, let us first recall that in the original Dicke model, $J_z=\frac{1}{2}\sum_{n=1}^N\sigma^z_n$ is the sum of $N$ individual (pseudo)-spin-$\frac{1}{2}$ operators. Superradiant states with maximal spin polarization are then energetically favored by $\mathcal{H}_N$, but there are only two configurations (all spins pointing either up or down) where that is achieved.  All other spin configurations  have larger (non-zero) entropy  such that  thermal fluctuations trigger a phase transition to a thermally disordered (normal) phase   above a critical temperature.

On the other hand, if the state space is restricted to the highly symmetric Dicke states $\ket{J, M}$, the configuration space is much smaller and there is no gain in entropy for states with higher energy. In this situation, which corresponds to $N$ bosons occupying either the upper or lower level of a two-level system, a thermal phase transition does not occur.

{\em The partition function and the number of configurations .--} We start with discussing the degeneracy factors that appear when evaluating the 
partition sum $Z_N$. Let us assume a configuration space with $N$ particles distributed among $N_s$ two-level `sites' $i$, all of which have the same  up-level $\uparrow$ and down-level $\downarrow$ energies.  We write $J_z \equiv \frac{1}{2}\sum_{i=1}^{N_s}\left(n_{i\uparrow}-  n_{i\downarrow}\right)$ with number operators $ n_{i\uparrow}$, $ n_{i\downarrow}$ such that
\begin{align}
Z_N = \sum_{n=0}^N c_n e^{\beta \frac{g^2}{N\omega}(N-2n)^2},
\end{align}
where $c_n$ is the number of configurations with $n$ particles in the down levels $\downarrow$ and $N-n$ in the up levels $\uparrow$. 

{\em Fermions .--} For fermions, 
 the number of configurations for $n$ particles in the down levels is $\binom{N_s}{n}$, and for the remaiming $N-n$ particles in the up levels it is $\binom{N_s}{N-n}$;
consequently $c_n= \binom{N_s}{n}\binom{N_s}{N-n}$. ii) For fermions with no two particles on the same site $i$, one has a restricted choice once all the $\uparrow$ (or the $\downarrow$) are occupied and thus $c_n= \binom{N_s}{n}\binom{N_s-n}{N-n}$. 
For the particular case $N_s=N$, this corresponds to the $N$ localized and {\em distinguishable} (pseudo) spins in the original Dicke model discussed above and gives the partition sum 
\begin{align}\label{ZD}
Z_N^{\rm D}\equiv \sum_{n=0}^N \binom{N}{n}e^{\beta \frac{g^2}{N\omega}(N-2n)^2}.
\end{align}

{\em Bosons .--}  For $N$  bosons in $m$ single particle levels, there are $\binom{N+m-1}{m-1}$ configurations which in our case means 
\begin{align}\label{cn}
c_n \equiv \binom{n+N_s-1}{N_s-1}\binom{N-n+N_s-1}{N_s-1}.
\end{align}
The particular case of a single site  $N_s=1$ yields $c_n=1$ and thus
\begin{align}\label{Zbos}
Z_N^{{\rm bos}}\equiv \sum_{n=0}^N e^{\beta \frac{g^2}{N\omega}(N-2n)^2},\quad N_s=1.
\end{align}
We also obtain this result by representing $J_z = \frac{1}{2}(2 b^\dagger b -N)$ via one Holstein-Primakoff boson $b^\dagger$, or alternatively by writing 
$J_z=\frac{1}{2}(b^\dagger_\uparrow b_\uparrow - b^\dagger_\downarrow b_\downarrow)$ with two Schwinger boson modes for $\uparrow$ and $\downarrow$ by using $b^\dagger_\uparrow b_\uparrow + b^\dagger_\downarrow b_\downarrow=N$ and noticing that the number $n$ of $\downarrow$-bosons uniquely fixes a configuration which means $c_n=1$.

{\em Landau free energy function .--} The most transparent way to discuss the difference between the two cases \eq{ZD} and \eq{Zbos} and generalizations thereof is by transforming the partition sums into integrals over an order parameter $y$. As we are dealing with a zero-dimensional field theory here, this is particularly simple and is formally achieved by a Hubbard-Stratonovich transformation which here is simply given by the Gaussian integral identity, $e^{x^2}=\frac{1}{\sqrt{\pi}}\int_{-\infty}^{\infty}dye^{-y^2+2xy}$, applied to the respective Boltzmann factors in $Z_N$. For the distinguishable (Dicke) case, we use the binomial formula as $\sum_{n=0}^N\binom{N}{n}e^{z(N-2n)}=(2\cosh z)^N$ to obtain (after substituting $y\to \sqrt{N} y$)
\begin{align}\label{phiD}
Z_N^{\rm D}&= \sqrt{\frac{N}{\pi}}\int_{-\infty}^{\infty}dye^{-N\Phi^{\rm D}(y)}\nonumber\\
\Phi^{\rm D}(y)&\equiv y^2-\ln\left(2\cosh\left(2\alpha y\right)\right),\quad \alpha\equiv g\sqrt{\frac{\beta}{\omega}},
\end{align}
where we introduced the dimensionless coupling parameter $\alpha$. 

Similarly, for the $N_s=1$ boson case, we carry out the geometric progression $\sum_{n=0}^N e^{z(N-2n)}=  e^{-Nz}(e^{2z(N+1)}-1)/(e^{2z}-1)$ to obtain
 \begin{align}\label{phiboson}
Z_N^{\rm boson}&= \sqrt{\frac{N}{\pi}}\int_{-\infty}^{\infty}dye^{-N\Phi^{\rm boson}(y)}\\
\Phi_N^{\rm boson}(y)&\equiv y^2+   2\alpha y-\frac{1}{N}\ln
\frac{e^{4\alpha (N+1) y }-1}{e^{4\alpha y}-1},\quad N_s=1.\nonumber
\end{align}
The so defined Landau free energy functions $\Phi(y)$ now allow us to elucidate the critical properties of the models.

\begin{figure}[t]
\includegraphics[width=\columnwidth]{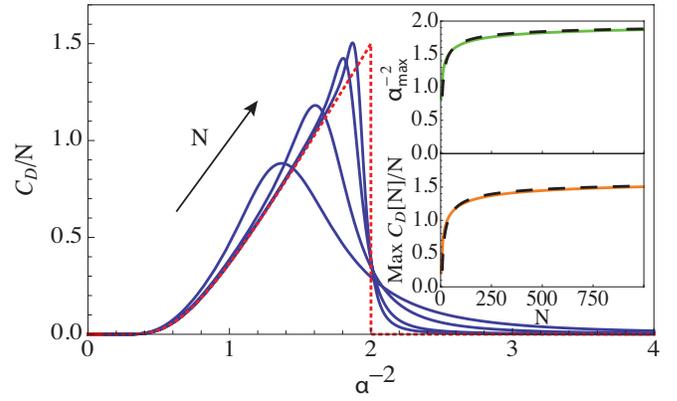}
\caption[]{\label{cdicke-over-n}The specific heat per particle ($k_B=1$)  as a function of $\alpha^{-2}\equiv \omega/( g^2\beta)$ (proportional to the temperature) in the Dicke case of $N$ localised spin--$\frac{1}{2}$s, (derived from the fermionic partition sum  \eq{ZD}), for $N=20,75,400$ and $1000$ particles (dotted line:  $N\to\infty$). 
Upper inset:  shift of peak from critical point $\alpha^*=1/\sqrt{2}$ for all $N\leq1000$. Lower inset: peak  height at maximum. Dashed lines: fit of the numerical data with $f_{upper}^{1}\equiv 1.64-\frac{3.9}{\sqrt{N}}$ and $f_{lower}^{1}\equiv 1.98-\frac{3.1}{\sqrt{N}}$.}
\end{figure}

First, we observe that in the (Dicke) case of distinguishable particles, $\Phi^{\rm D}(y)$ in \eq{phiD} is $N$-independent, and $\Phi^{\rm D}(y) \equiv \beta f$
is determined by the $\omega_{0}=0$--limit of the usual mean-field ($N \to\infty$) expression for the free energy $f$ per particle \cite{WH73,HL73a,CGW73}.
As expected, the term  $\ln\left(2\cosh\left(2\alpha y\right)\right)$ therefore is the mean field free energy of a single (pseudo) spin in the fluctuating field $y$. The parameter $\alpha$ determines the shape of $\Phi^{\rm D}(y)$ and, in the $N\to \infty$ limit, the position $y_{\rm D}$  of the minimum of $\Phi^{\rm D}(y)$ relevant for the asymptotic expression of the integral according to the Laplace method, as given by the self-consistent equation
\begin{equation}
y_{\rm D} = \alpha \tanh 2\alpha y_{\rm D}.\label{selfcon-eq}
\end{equation}
This has the unique solution $y_{\rm D}=0$ when $2\alpha^2<1$ corresponding to temperatures $T>T_c\equiv 2g^2/\omega$ larger than the critical temperature $T_c$. This solution describes the normal phase whereas for $T<T_c$ there exist two minima in $\Phi^{\rm D}(y)$ that describe the symmetry-broken superradiant phase.

This is in contrast to the bosonic case, where the Landau free energy $\Phi_N^{\rm boson}(y)$ in \eq{phiboson} is $N$-dependent but acquires a simple form in the thermodynamic limit $N\to \infty$
\begin{align}\label{phibosoninfty}
\Phi_\infty^{\rm boson}(y) = y^2 -2\alpha |y|,
\end{align}
which has two unique minima at $y=\pm \alpha$ regardless of the value of $\alpha$. This means that in the bosonic case, one is always in the superradiant, symmetry-broken phase and no thermal phase transition  into a normal phase occurs.

{\em Thermodynamic behavior.--}
In the following, we discuss the specific heat $C\equiv\beta^{2} \partial^{2}_{\beta} \log(Z_{N})$  for both cases (we set $k_B=1$).
In the Dicke case (spin--$\frac{1}{2}$s corresponding to the fermionic partition sum \eq{ZD}), Fig. (\ref{cdicke-over-n}),  the specific heat per particle can be calculated from $\Phi^{\rm D}(y) \equiv \beta f$ in the thermodynamic limit $N\to \infty$ by eliminating derivatives of $y_D$ and solving 
the self-consistent \eq{selfcon-eq}, leading to the expected singular behaviour at the transition point $\alpha^*=1/\sqrt{2}$.
A $1/N$ expansion of $ Z_N^{\rm D}$, \eq{phiD},  works only well not too close to the critical point, as we have checked. Numerically, for finite $N$ the peak height of $C/N$ and the shift of the peak from $\alpha^*$ give, however,  a very good agreement with a $ {1}/\sqrt{N}$ correction fit.

\begin{figure}[t]
\includegraphics[width=\columnwidth]{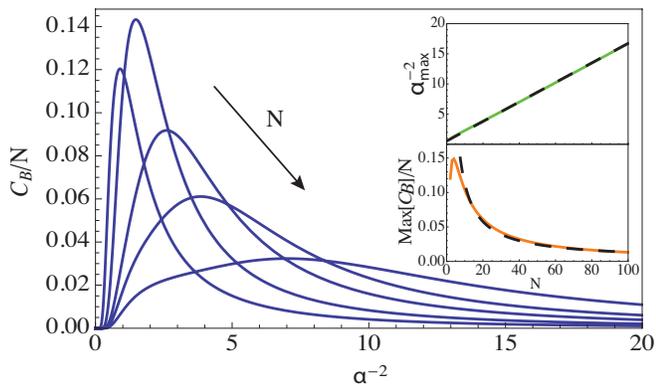}
\caption[]{\label{cind-over-n}The specific heat per particle as in Fig. (\ref{cdicke-over-n}) but for the bosonic partition sum   \eq{Zbos}, for $N=2,5,12,20$ and $40$ particles. 
Dashed lines: linear fit $f_{upper}^{2}\equiv 0.6+0.2 N$ for maximum position (upper inset), $f_{lower}^{2}\equiv \frac{1.1}{N}$ fit for maximum (lower inset). }
\end{figure}

In the bosonic case, Fig.  (\ref{cind-over-n}), the specific heat per particle shows a totally  different behavior. 
For small numbers of bosons ($N=5,6,...12$),   ${C}/{N}$ first increases with $N$ and then (for $N > 12$) decreases with $\frac{1}{N}$, as does its maximum. The position $\alpha^{-2}$ of the maximum of ${C}/{N}$ is linear the particle number $N$.  In the limit $N \to \infty$ the specific heat per particle is zero which can be understood from \eq{phibosoninfty} by using the above mentioned thermodynamic relations and considering that $y=\pm\alpha$.

\begin{figure}[t]
\includegraphics[width=\columnwidth]{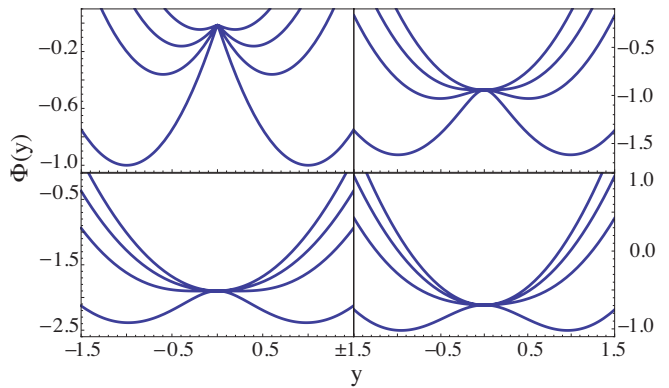}
\caption[]{\label{figure1}Crossover between bosonic and Dicke case in the 
Landau free energy, \eq{landaugen} for $N=400$ bosons and coupling strengths $\alpha=0.2,0.4,0.6,1.0$ within each graph. Upper left $N_s=1$  ($N_s$: number of two-level sites), this corresponds to the simple bosonic case \eq{phiboson} (and \eq{phibosoninfty} for $N\to \infty$), i.e. the absence of a phase transition, since the number of minima remains two irrespective of $\alpha$. The phase transition is re-established for macroscopic degeneracy (upper right $N_s=N/4$, lower left $N_s=N$), since a transition from a double- to single-minimum occurs. The lower right shows the Dicke case $\Phi^{\rm D}(y)$, \eq{phiD}, with the phase transition occurring at $\alpha^* = 1/\sqrt{2}\approx 0.71$.}
\end{figure}

{\em Crossover between bosonic and Dicke case .--} We now discuss an interesting crossover between the two cases obtained above by regarding the bosonic case with $N_s>1$. For $N_s$ of the order of the particle number $N$, we expect the bosons to spread over many energetically equivalent configurations which, due to the entropy argument given in the introduction, should re-establish the thermal phase transition found in the Dicke (fermionic spin--$\frac{1}{2}$) case. We therefore generalize  the Landau free energy for the bosonic case \eq{phiboson} to arbitrary $N_s$,
\begin{align}\label{landaugen}
&\Phi_{N,N_S}^{\rm boson}(y)\equiv y^2-\frac{1}{N}\ln \sum_{n=0}^N c_n   e^{2\alpha y(N-2n)}    
\end{align}
with $c_n$ given by \eq{cn}.
Fig. (\ref{figure1}) shows the crossover in the free energy when passing from the bosonic case with $N_s=1$ (showing no phase transition when varying $\alpha$) to larger degeneracies, where for $N_s\gg N$, the free energy  $\Phi_{N,N_S}^{\rm boson}(y)$ becomes equivalent to the one of the Dicke case, \eq{phiD}. 
This can be understood by using Stirling's formula to expand the number of configurations $c_n$, \eq{cn},  which leads to $c_n \sim \binom{N}{n} (N_s/N)^N$ and therefore the partition sum
\begin{align}
Z_N^{{\rm bos}} \sim \left( \frac{N_s}{N} \right)^N Z_N^{{\rm D}},\quad  N_s\gg N\gg 1.
\end{align}
Correspondingly, in this limit the bosonic free energy per particle differs from the Dicke free energy by $-k_B T  \ln (N_s/N)$ which just describes an additional entropy term caused by the enhanced 
`volume' of configurations. Thus for $N_s\gg N$, multiply occupied sites play no r\^{o}le any longer, nor does the statistics of the particles (fermions or bosons), and the only remaining relevant statistical 
information is the number of occupied upper and lower levels as in the Dicke case. 
As the entropy gain $k_B \ln (N_s/N) $ is just a constant, the specific heat $C/N$ then has to coincide with the specific heat in the Dicke case. 

\begin{figure}[t]
\includegraphics[width=\columnwidth]{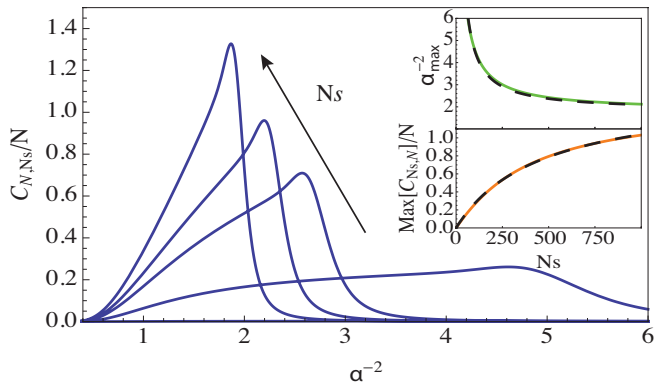}
\caption[]{\label{cv-crossover}The specific heat per particle at fixed particle number $N=400$ for different number of two-level sites $N_{s}=1$ (bosonic case, bottom line), $N_{s}=100$, $N_s=400$ (Dicke case)  and $N_s =600,5000$ as obtained from \eq{z-general}.  
Upper inset: with increasing $N_{s}$ the peak position in $C/N$ shifts towards the critical point $\alpha^{-2}=2$ of the Dicke case. Lower inset: 
increase of maximum of $C/N$ with $N_s$. Dashed lines: fit of the numerical data with $f_{upper}^{4}\equiv 1.82-\frac{275}{N_s}$ and $f_{lower}^{4}\equiv \frac{1.5 N_s}{458.6 + N_s}$. }
\end{figure}

This behavior of  the specific heat $C/N$ is shown in Fig. (\ref{cv-crossover}), 
where  we vary  $N_{s}$ from the bosonic case $N_s =1$ over  $N_s = N$ to finally higher values of $N_{s} \gg N$ corresponding to the Dicke (fermionic) form of  $C/N$ from Fig. (\ref{cdicke-over-n}).  
With $N$ fixed, the $C/N$ curves have peaks that shift  from larger to smaller values of $\alpha^{-2}$ with increasing $N_{s}/N$ 
corresponding to a {\em decreasing} $N/N_s$ in the $N_s=1$ bosonic case in Fig. (\ref{cind-over-n}). 
At even larger $N_s/N \gg 1$ the curves approach the  Dicke form at the critical point $\alpha^*$ of the Dicke thermal phase transition.
Note that the peaks approach $\alpha^*$ from the right here, whereas  in the Dicke case, Fig. (\ref{cdicke-over-n}), $N_s=N$ was fixed and  the   $C/N$ peaks approached $\alpha^*$
from {\em smaller} $\alpha^{-2}$ with increasing $N$.

{\em Conclusion and outlook .--} 
Our results indicate that  the Dicke model displays an interesting thermodynamic behavior if one  considers  the possible  configurations of $N$ particles in $N_s$ two-level systems. 
In the ultra-strong coupling limit, the model reduces to an atomic self-interaction term for which we have derived explicit expressions for  the partition function, Landau free energy and specific heat.
We find a crossover in the specific heat from a bosonic form at small $N_s$ (where no thermal phase transition occurs)  to the form following from  the (original) Dicke model for  $N_s\gg N \gg 1$ that displays a thermal phase transition at $\alpha^{-2}\equiv    k_BT \omega/ g^2 = 2$ between the normal and the superradiant phase. 

In our calculations, we regarded $ \mathcal{H}_N$ in \eq{z-general} as an effective  Hamiltonian for the ultra-strong coupling regime between atoms and light of the Dicke model \eq{HDicke}. In analogy with  the simulation of the Dicke model with Bose-Einstein condensates \cite{BGBE10}, we suggest the various limits of $ \mathcal{H}_N$ to be regarded as effective models to be simulated with, e.g., cold atoms. 
A further challenge would be an extension to Dicke-type models with finite coupling strengths, multiple levels \cite{HEB11}, and degeneracies $N_s$ in the bosonic case.

We acknowledge support by the DFG through SFB 910.  M.A.A. thanks M. Hayn, V. Bastidas, B. Pimentel and N. Svaiter for useful discussions, and the Institut f\"ur Theoretische Physik of the TU Berlin for their kind hospitality. M.A.A. acknowledges FAPESP for financial support.


\begin{thebibliography}{10}

\bibitem{Dic54}
R.~H. Dicke, Phys. Rev. {\bf 93},  99  (1954).

\bibitem{Bra05}
{T. Brandes}, Physics Reports {\bf 408/5-6},  {315:474}  (2005).

\bibitem{EB03two}
{C. Emary and T. Brandes, Phys. Rev. Lett. {\bf 90}, 044101 (2003); Phys. Rev.
  E 67, 066203} (2003).

\bibitem{AH12}
{A.~Altland and F.~Haake, arXiv:1110.1270, 1201.6514} (2012).

\bibitem{VD06}
{J. Vidal and S. Dusuel, Europhys. Lett. \textbf{74}, 817 } (2006).

\bibitem{HL73}
K. Hepp and E. Lieb, Ann. Phys. {\bf 76},  360  (1973).

\bibitem{PhysRevA.75.013804}
F. Dimer, B. Estienne, A.~S. Parkins, and H.~J. Carmichael, Phys. Rev. A {\bf
  75},  013804  (2007).

\bibitem{BGBE10}
{K. Baumann, C. Guerlin, F. Brennecke, and T. Esslinger}, nature {\bf 464},
  1301  (2010).

\bibitem{LEB04}
{N. Lambert, C. Emary, and T. Brandes}, Phys. Rev. Lett. {\bf 92},  073602  (2004).

\bibitem{WH73}
{Y. K. Wang and F. T. Hioe}, Phys. Rev. A {\bf 7},  831  (1973).

\bibitem{HL73a}
K. Hepp and E. Lieb, Phys. Rev. A {\bf 8},  2517  (1973).

\bibitem{CGW73}
{H. J. Carmichael, C. W. Gardiner, and D. F. Walls}, Phys. Lett. {\bf 46A},  47
   (1973).

\bibitem{ALS07}
{M. Aparicio Alcalde and A. L. L. de Lemos and N. F. Svaiter} Journ. Phys. A \textbf{40}, 11961 (2007); M. Aparicio Alcalde and B. M. Pimentel, Physica A {\bf 390}, 3385 (2011).

\bibitem{HEB11}
{M. Hayn, C. Emary, and T. Brandes}, Phys. Rev. A {\bf 84},  053856  (2011).

\end{thebibliography}
\end{document}